# Dosimetric commissioning and quality assurance of scanned ion beams at the Italian National Center for Oncological Hadrontherapy

Alfredo Mirandola[1], S. Molinelli[1], G. Vilches Freixas[1], M. Donetti[1,2], A. Mairani[1], E. Gallio[1], D. Panizza[1], S. Russo[1], G. Magro[1,3], S. Giordanengo[2] and M. Ciocca[1] R. Orecchia[1,4]

1. Fondazione CNAO, strada Campeggi 53, Pavia 27100, Italy and Radiotherapy Division,
2. INFN, Torino 10125, Italy
3. Università degli Studi di Pavia, Via U. Bassi, Pavia 27100
4. European Institute of Oncology, Via Ripamonti 435, Milano 20141, Italy

**Purpose:** To describe the dosimetric commissioning and quality assurance (QA) of the actively scanned proton and carbon ion beams at the Italian National Center for Oncological Hadrontherapy.
**Methods:** The laterally integrated depth-dose-distributions (IDDs) were acquired with the PTW Peakfinder, a variable depth water column, equipped with two Bragg peak ionization chambers.
FLUKA MC code was used to generate the energy libraries, the IDDs in water, and the fragment spectra for carbon beams. EBT3 films were used for spot size measurements, beam position over the scan field, and homogeneity in 2D-fields. Beam monitor calibration was performed in terms of number of particles per MU using both a Farmer-type and an Advanced Markus ionization chamber. The beam position at the isocenter, beam monitor calibration curve, dose constancy in the center of the spread-out-Bragg-peak, dose homogeneity in 2D-fields, beam energy, spot size, and spot position over the scan field are all checked on a daily basis for both protons and carbon ions and on all beam lines.
**Results:** The simulated IDDs showed an excellent agreement with the measured experimental curves. The measured FWHM of the pencil beam in air at the isocenter was from 0.7 to 2.2 cm for protons. For carbon ions, two sets of spot size are available: from 0.4 to 0.8 cm (for the smaller) and from 0.8 to 1.1 cm (for the larger one).
The spot position was within ± 1 mm over the whole 20x 20 cm$^2$ scan field; homogeneity in a uniform squared field was within ± 5% for both particle types at any energy. QA results exceeding tolerance levels were rarely found. In the reporting period, the machine downtime was around 6%, of which 4.5% was due to planned maintenance shutdowns.
**Conclusions:** After successful dosimetric beam commissioning, quality assurance measurements performed during a 24-month period show very stable beam characteristics.

## 1. INTRODUCTION

Light ion radiotherapy is a rapidly developing technique and is becoming an important treatment modality within the field of radiotherapy. Due to the encouraging physical and biological characteristics of protons and carbon ions, several facilities are being built worldwide and new projects have been proposed in the last few years.[1]
Light ion beams can be delivered using different approaches, including the passive technique, uniform scanning, and pencil beam scanning (spot scanning or raster scanning).[2,3] The National Center for Oncological Hadrontherapy (CNAO) is the first hospital-based light ion therapy facility built in Italy. It is equipped with an "in-house" built synchrotron and beam monitor (BM) systems, providing actively scanned proton and carbon ion pencil beams. The scanning magnets are positioned approximately 5.5 m upstream from the isocenter and therefore provide a small beam divergence.
As a consequence of this, the beams delivered at CNAO can be classified as *quasiparallel* scanned beams. In treatment planning, the target volume is dissected into isoenergy layers and the dose is delivered by spot-by-spot active scanning. The beam energy is selected using the synchrotron and then each layer is scanned with the discrete energy requested to deposit dose at the desired depth in tissue. For each energy layer, a pair of magnets rapidly scans the beam in the transverse plane until all spots within the layer are irradiated, before moving to the next isoenergetic layer. Four discrete levels of beam intensity ($I$) can be selected: $I_{max}$ and 50%, 20%, 10% of $I_{max}$. $I_{max}$ is approximately $3.0 \times 10^{+09}$ and $6.0 \times 10^{+07}$ particles per second for protons and carbon ions, respectively.

At CNAO, three treatment rooms are available: two of the rooms are equipped with a single horizontal fixed beam line, while in the third one, a horizontal and vertical fixed beam line[4] is available. Each beam line is equipped with two independent beam monitors: Monitor1 and Monitor2, positioned in the beam nozzle (Fig. 1). Monitor1 is positioned at a larger distance from the isocenter and consists of a large transmission ionization chamber that intercepts the entire beam and two orthogonally oriented strip chambers for spot position measurements in the transverse plane. Monitor2 is positioned closest to the isocenter and consists of another transmission ionization chamber and a pixel chamber for 2D spot position measurements.[5] The CE-marked and commercial treatment planning system (TPS) Syngo RT (Siemens AG Healthcare, Erlangen, Germany) is used. It supports the dose optimization and calculation for both active scanned protons and carbon ions. Once approved, treatment plans are transferred to the Mosaiq Oncology Information System (OIS) (Elekta Medical Systems, Sunnyvale, CA) and are finally sent to the treatment console for dose delivery.

Patients are positioned in one of the treatment rooms using the 6 degrees of freedom patient positioning system (PPS);[6,7] pretreatment imaging in the treatment rooms equipped with the horizontal beam line is performed using the patient verification system (PVS), consisting of two pairs of ceiling-mounted x-ray tubes and amorphous-silicon flat panels. In the third room, a six-joint robot carrying a C-arm equipped with an x-ray tube and a flat panel is installed.

During the preclinical phase, the physical and biological[8] characteristics of proton and carbon ion beams were commissioned. In this report, the commissioning of the physical and dosimetric characteristics is described. The CNAO procedures, instrumentations, and measurements to ensure accurate and safe treatments, together with the results of 24 months of daily quality assurance (QA) tests, are presented here.

The results of the patient-specific pretreatment plan dosimetric verification have been published previously.[9]

## 2. MATERIALS AND METHODS

The main requirements for the treatment beam specifications were carefully predefined by a multidisciplinary panel of experts, including radiation oncologists, medical, and accelerator physicists, taking into account the expected tumor types to be treated. Specifications were given in terms of particle type, energy range in water, depth-spacing between energies, spot size, beam intensity, and field dimensions. The careful selection of these parameters was especially important, as CNAO was designed to be a general purpose light ion radiotherapy facility.

Due to the complexity of the diseases to be treated, as well as the highly innovative technology employed and the physical and dosimetric issues related to the use of proton and carbon ion pencil beams, strict commissioning and periodic QA programs were established. An extensive set of measurements was performed in order to ensure high levels of quality and safety for patient treatments. Standard methods for the commissioning and QA of both proton and carbon ion beams have not been established yet. However, for QA of proton beams, the recommendations given in the ICRU 78 report[10] can be adopted. At CNAO, such procedures were developed with reference to the most relevant literature available[11–15] and by following the guidelines stated by the TPS vendor, Siemens AG.[16] During the commissioning phase, the values of all functional performance characteristics were determined. These results also served as a reference for all subsequent QA tests, which are performed periodically in order to detect any change in these parameters.

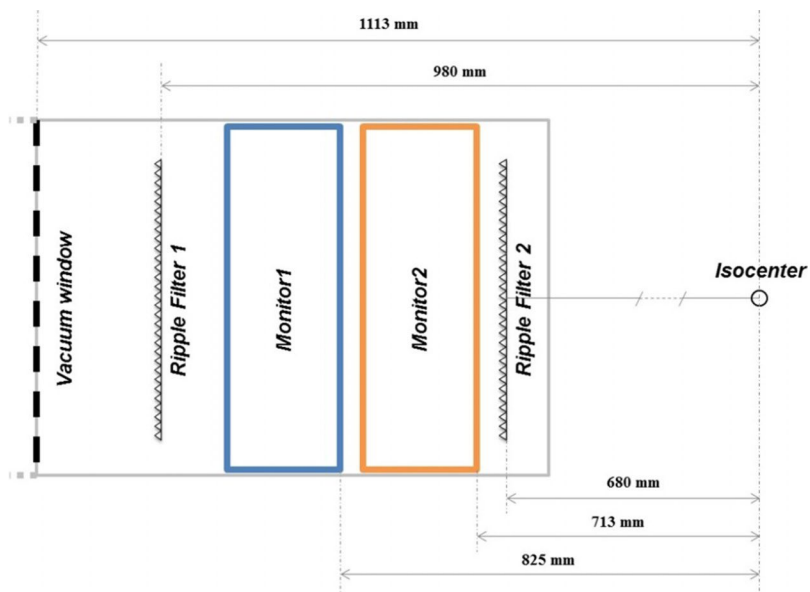

Figure. 1. A schematic drawing of the beam line showing the relative distances from the isocenter.

## 2.A. Beam commissioning

In the following paragraphs, the physical and dosimetric characteristics of the particle beams, assessed during commissioning, are reported. Additionally, one of the most critical issues in light ion radiotherapy, the relation between particle ranges relative to water and CT-numbers [CT-Hounsfield units (HU)] is described. A schematic summary of the performed procedures is reported in Table I. The remaining fundamental topics to be managed during commissioning, such as the patient positioning and x-ray verification system and interlock acceptance and testing, are not explicitly reported in this paper as they have already been described by Giordanengo *et al.*,[5] Pella *et al.*,[6] and Desplanques *et al.*[7]

### 2.A.1. Energy library and depth dose distributions

The energies commissioned at CNAO range from 62.73 to 228.57 MeV for proton beams and from 115.23 to 398.84 MeV/u for carbon ions. The corresponding depths of Bragg peaks in water extend from 30 to 320 mm for protons and from 30 to 270 mm for carbon ions, with steps of 2 mm for both particle types. In order to treat tumors at depths shallower than 30 mm, *range shifters* of different thicknesses consisting of slabs of RW3 (PTW, Freiburg, Germany) joined together, are used. The FLUKA Monte Carlo (MC)[17–20] code was adopted to generate the list of energies to be provided by the synchrotron within the allowed range, similarly to Parodi *et al.*[21] The integrated depth-dose-distributions (IDDs) were measured with the Peakfinder (PTW, Freiburg, Germany), a variable depth water column controlled by dedicated acquisition software and equipped with two plane-parallel ionization chambers, each of 4.08 cm radius (PTW Bragg peak ionization chambers TM34080 as field chamber, and PTW TM34082 as reference chamber). The ionization chambers were connected to the PTW Tandem dual electrometer. The IDDs were calculated as the ratio between the readings of the two chambers. The field detector can be moved in the water column with an extremely high spatial resolution, down to 10 $\mu$m steps for very precise data acquisition in the Bragg peak region. The agreement between IDDs calculated using FLUKA and those measured in water was carefully assessed during the beam commissioning. Data acquisition with the Peakfinder was performed for each particle type and beam line. The pristine IDDs for protons were acquired with and without a ripple filter (2 mm thickness), while the IDDs for carbon ions were acquired without ripple filters and with two crossed ripple filters (each of 2 mm thickness).[22] For the first two fixed horizontal lines commissioned (one room for protons and one room for carbon ions), each of the 267 energies available (146 for protons and 121 for carbon ions) was measured and used as a benchmark for the FLUKA MC code. The measured IDDs were compared to MC simulations performed by scoring deposited energy in a cylinder of 4.08 cm radius (corresponding to the radius of the Bragg-peak chamber). However, the size of the Bragg peak chamber is not large enough to capture the whole laterally scattered dose, especially for the low energy beams.[12,23,24] Therefore, in order to take into account the full lateral scatter, as well as the dose from nuclear fragmentation, an additional set of MC simulations, in which the deposited energies were scored in a larger area ($25 \times 25$ cm$^2$), was performed and used as the input for the TPS (see Subsection 3.A). For the commissioning of the other beam lines, a representative subset of energies covering the whole range (about 10% of the available energies) were measured and compared with those already accepted. The remaining IDDs were then simulated by MC.

Table I. Summary of the program of the main measurements needed for beam commissioning.

| Test type | Beam type | Method |
|---|---|---|
| Beam energy (cf. Sec. 2.A.1) | Monoenergetic pencil beam | Peakfinder |
| Beam contamination (cf. Sec. 2.A.2) | Monoenergetic pencil beam | Peakfinder |
| Spot size in air (cf. Sec. 2.A.3) | Monoenergetic pencil beam | EBT3 films + film holder |
| Spot position (isocenter) (cf. Sec. 2.A.4) | Monoenergetic pencil beam | EBT3 films |
| Spot position over the scan field (cf. Sec. 2.A.4) | Monoenergetic scanned pencil beam | EBT3 films + All-in-one phantom |
| Dose homogeneity in 2D-fields (cf. Sec. 2.A.5) | Monoenergetic, uniform scanning beam | EBT3 films + film holder |
| BM homogeneity and calibration (cf. Sec. 2.A.7) | Monoenergetic, uniform scanning beam | Farmer IC/Advanced Markus + water phantom |
| Dose in a reference volume (cubic volumes) (cf. Sec. 2.A.8) | Modulated scanning beam, SOBP | Farmer IC/Advanced Markus + water phantom |
| Dose distributions in water phantom (cf. Sec. 2.A.8) | Modulated scanning beam, SOBP | PinPoint chambers + 3D holder in water phantom |
| Dose distributions in inhomogeneous phantom (cf. Sec. 2.A.9) | Modulated scanning beam, SOBP | PinPoint chambers + inhomogeneous phantom |
| CT-range calibration (cf. Sec. 2.A.9) | Monoenergetic pencil beam | CT scan + Peakfinder |

### 2.A.2. Beam contamination

To verify possible beam contaminations, the Bragg curves were analyzed to assess the absence of secondary peaks apart from the Bragg peak corresponding to the selected energy. No beam contaminations were found in the commissioned energy range for both protons and carbon ions.

### 2.A.3. Spot size in air

The spot sizes in air were measured in terms of full width half maximum (FWHM). For that purpose, calibrated EBT3 films (International Speciality Products, Wayne, NJ) were irradiated with an unscanned pencil beam at isocenter position. The beam optics were tuned to extract two different energy- dependent sets of spot sizes for carbon ions and one set for protons. For carbon ions, a bigger spot size can be advantageous especially to treat very large tumors or moving targets more efficiently.

The spot sizes were measured with and without ripple filters inserted for both particle species.

http://dx.doi.org/10.1118/1.4928397

Irradiated films were scanned using an EPSON 10000 XL flatbed scanner and $FWHM_{x,y}$ were acquired. The analysis of the transversal profiles in terms of relative dose was performed using the PTW MEPHYSTO dedicated software. The applied spatial resolution for spot size determination was equal to 0.2 mm (150 dpi acquisition).

For the evaluation of the pencil beam widening, the spot size in air was measured at different distances from the vacuum window:

(a) For protons, five different positions were chosen: at the exit window, at Monitor2, at the isocenter, at 35 cm before, and 50 cm beyond the isocenter (the distance from the vacuum window to the isocenter is 1113 mm). For the first treatment line commissioned, spot sizes for all 146 proton energies were measured at isocenter, while spot sizes for 10 representative energies were measured at the other positions. For the following pro- ton beam line commissioned, a subset of 25 energies was measured at isocenter position.

(b) For carbon ions, three positions were chosen: at Monitor2, at the isocenter, and 50 cm beyond the isocenter. For every beam line, the spot sizes for 25 representative energies were measured at isocenter, while spot sizes for a subset of 10 energies were measured at the other positions.

The collected data were then implemented into the TPS. The same measurements, performed for all the beam lines, showed similar behavior in terms of FWHM.

### 2.A.4. *Spot position at the isocenter and across the scan field*

In order to check the accuracy of the pencil beam deflection by the scanning magnets across the whole scanning area ($20 \times 20$ cm$^2$), a spot pattern with nine beam spots in nominal positions (8 peripheral + 1 central) was created. This spot pattern was then delivered to the dedicated All-in-one PTW phantom (Fig. 2). The center of the phantom, loaded with EBT3 films, was aligned at the isocenter using the room lasers. Pencil beams of various energies were deflected toward the nominal positions, where steel spheres (diameter of 2 mm) are embedded in the phantom. The analysis of the transversal profiles was then performed using MEPHYSTO software with a spatial resolution of 0.2 mm.

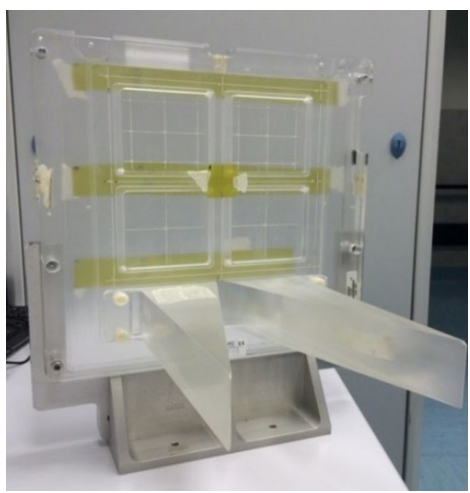

Figure. 2. All-in-one phantom loaded with strips of an EBT3 film for spot position measurements.

### 2.A.5. *Dose homogeneity in 2D-fields*

The 2D-field homogeneity was evaluated using EBT3 films placed in air at the isocenter. The films were irradiated with square monoenergetic uniform fields (i.e., constant number of particles delivered per spot) of different sizes (typically 6× 6 cm$^2$ and 15× 15 cm$^2$). The two orthogonal and diagonal profiles were evaluated and the homogeneity ($H$) was calculated as follows:

$$H = 100 \times \frac{D_{max} - D_{min}}{D_{max} + D_{min}}, \qquad (1)$$

where $D_{max}$ and $D_{min}$ are the maximum and minimum values of the dose along the selected profile. The homogeneity was evaluated within 80% of the 15× 15 cm$^2$ and within 70% of the 6× 6 cm$^2$ field size and diagonal profiles. During commissioning, the homogeneity in the 2D-fields was analyzed for every beam line and for both particle species available, at various energies, beam intensities (number of particles per second), and dose levels.

### 2.A.6. Determination of absorbed dose to water under reference conditions

Absorbed dose to water for scanned particle beams was determined by following IAEA TRS-398 Code of Practice,[25] adapted using the formalism specifically introduced by Hartmann et al.[13–15] for scanned particle beams. A square monoenergetic homogeneous field (6× 6 cm$^2$, both for protons and carbon ions) was delivered and dose was measured with a Farmer-type (PTW type 30013) and, for comparison, an Advanced Markus chamber (PTW type 34045) placed in a PTW water phantom type 41023 at the isocenter, at the fixed depth of 2 cm and connected to the PTW Unidos-webline electrometer. The dose was measured for a subset of representing energies in the commissioned range for both particle species. The depth of 2 cm in water does not fulfill the conditions of initial plateau region for all the available energies (at least for beam energies with Bragg peak ranging from 30 to 50 mm in water), and secondary particles can be produced for carbon ions at that depth in water. However, to take into account these aspects, the averaged mass stopping power of particles with initial energy $E$ at the depth of measurement $z$, as calculated using FLUKA MC code, was adopted [Eq. (5)].
In addition, to take into account the effective point of measurement of the cylindrical Farmer chamber, the measure- ment position was displaced by an additional depth of 2.3 mm (0.75 its inner radius), as showed by Jäkel et al.[26] and Palmans and Verhaegen.[27]

The square uniform fields were delivered using a trans- versal scanning step of 3 mm for protons and 2 mm for carbon ions at several representative beam energies.
Dose was determined based on the $N_{D,W}$ formalism,[25]

$$D_w = M \cdot N_{D,w} \cdot k_Q, \qquad (2)$$

where $M$ is the reading of the dosimeter corrected for deviations from reference conditions, $N_{D,w}$ is the chamber calibration factor, and $k_Q$ is the factor to correct for the difference between the response of that type of ionization chamber in the reference beam quality $Q_0$ used for chamber calibration and in the actual user beam quality $Q$. Some issues regarding the terms in the formula, in case of particle beams, are discussed in Secs. 2.A.6.a and 2.A.6.b.

*2.A.6.a. Correction of electrometer reading, M.* In addition to the corrections made due to pressure, temperature, and humidity[25] variations, dose corrections due to ion recombination and polarity effects were calculated for the reference Farmer chamber. Following the IAEA TRS-398 Code of Practice, the approximation to a continuous scanned beam in this case can be applied[13,28,29] as the spill duration of about 1 s is much longer than the chamber charge collection time. The ion recombination was evaluated and the corresponding correction factor ks was calculated for protons and carbon ions in the plateau region at two representative energies and several applied voltages from 100 to 400 V. Selected energies were 150.71 and 380.45 MeV/u for carbon ions and 81.56 and 198.11 MeV for protons.

The reciprocal charge was plotted over reciprocal voltage and fitted by a second-order polynomial. For the Farmer chamber, the linear term was found to be negligible[13] so the two voltage method formula can be adopted,

$$k_s = \frac{(V_1/V_2)^2 - 1}{(V_1/V_2)^2 - M_1/M_2}, \quad (3)$$

where $V_1$ is the operating voltage +400 V, $V_2$ the +133 V second voltage, while $M_1$ and $M_2$ are the corresponding obtained collected charges.

The correction factors calculated for protons were negligible, being equal to 1.003 for both the beam energies investigated, while it was 1.008 and 1.005 for the lower and higher carbon ion energy, respectively, which is in close agreement with Hartmann *et al.*[13]

Since both the uncertainty reported by the chamber manufacturer in the calibration certificate for the missing $k_p$ correction for the Farmer chamber under Co-60 beams (less than 0.2%) and the polarity effect for ions and protons ($k_p$ equal to 1.002 and 1.001, respectively) are negligible, the polarity effect can be neglected.

*2.A.6.b. Beam quality correction, $k_Q$.* The adopted beam quality correction factors are tabulated in IAEA TRS-398 (Ref. 25) and used in dose determination both for protons and carbon ions. For proton beams, $k_Q$ is calculated as a function of the beam quality index $R_{res}$,

$$R_{res} = R_p - d, \quad (4)$$

where $R_p$ is the practical range, defined as the depth distal to the Bragg peak at which the dose is 10% of its maximum value and $d$ is the depth of measurement in ICRU (Ref. 42) and IAEA.[25] This factor varies by approximately 0.3% as a function of $R_{res}$ and a tabulated $k_Q$ value of 1.029 was used for the Farmer chamber in this case.

For carbon ions, $k_Q$ is assumed to be independent of the beam quality, due to a current lack of experimental data.[25] In this case, the tabulated value of 1.032 was used in this case for the Farmer chamber.

Further considerations involving dose determination in case of carbon ion beams will be addressed in Sec. 4.

### 2.A.7. BM calibration

In order to calibrate the beam monitor chambers in terms of number of particles per monitor unit (MU), a series of square monoenergetic uniform fields were delivered, as described in Jäkel *et al.*[15] The calibration factor at a specific energy $E_i$ - $C(E_i)$ - is defined as the number of particles ($N_p$) per monitor unit and is calculated as follows:

$$C(E_i) = \frac{N_P}{MU} = D_{meas} \cdot \frac{\Delta x \cdot \Delta y}{MU \cdot S_{E_i(z)}},$$

where $D_{meas}$ is the absorbed dose measured in the phantom, $S_{E_i(z)}$ is the mass stopping power of particles with initial energy ($E_i$) at the depth of measurement $z$ (2 cm in water) as calculated using the FLUKA MC code. $\Delta x$ and $\Delta y$ are the spacing between two consecutive spots in the transversal direction (3 and 2 mm for protons and carbon ions, respectively). The delivered number of particles was set as a constant for each scan point, i.e., $1.0 \times 10^{+08}$ and $2.0 \times 10^{+06}$ particles per spot for protons and carbon ions, respectively. A representative set (typically 9) of individual $C(E_i)$ in the whole range of beam energies were acquired. A third order polynomial curve, $F(E)$, was adopted to fit the collected data and was used as BM calibration curve. The curve equation $F(E)$, for each beam line and particle type, was then hardcoded in the software managing the BM system and is used for treatments. During this phase, the proportionality and short-term reproducibility of the BM system, as well as its dependence on the beam intensity (i.e., number of particles per second), were investigated.

### 2.A.8. *Determination of absorbed dose to water in the center of the spread-out-Bragg-peak (SOBP)*

As the beam monitors are calibrated in the plateau region, the dose in the center of the SOBP is also measured to check that the superposition of the Bragg peaks and the resulting dose distribution are both correct. In order to check the whole set of available energies, four different cubes of $6 \times 6 \times 6$ cm$^3$, centered at depths of 9, 15, 21, and 24 cm, respectively, in water, were created in the TPS for both particle types. Each cube consists of 31 layers (i.e., energies), which are homogeneous along the lateral and orthogonal directions. As for the reference conditions described above, a pair of ripple filters is inserted for carbon ions. Doses were measured using both Farmer and Advanced Markus ionization chambers, placed at the center of the cube. Additionally, during commissioning and after every major TPS upgrade,[9] dose measurements in the center of the SOBPs were also performed with PTW PinPoint ionization chambers model T31015.

The dose dependency on the particle fluence was evaluated by delivering beams at four dose levels: 0.5, 1, 2, and 5 Gy. The dependency on beam intensity was also tested and doses were measured for the four available levels $I_{max}$ and 50%, 20%, 10% of $I_{max}$.

### 2.A.9. *CT-HU calibration*

The rationale of this topic is to convert the CT numbers to the water-equivalent path length (WEPL) of the materials under examination.[30] That is of fundamental importance for the correct calculation of the particle range within the patient. That procedure involved two steps: first, a CT scan of a CIRS electron density phantom (model 062, CIRS, Inc., Norfolk, VA), equipped with several tissue substitutes (lung inhale, lung exhale, adipose, breast, liver, muscle, trabecular bone, dense bone, water filled syringe, and titanium), was acquired using each of the planning CT protocols. The mean HU were sampled for each material. Second, the pristine IDDs in water as well as IDDs with probes of each material positioned in the beam path were acquired with the Peakfinder, as already described in Subsection 2.A.1. The relative WEPL (rWEPL) was determined as the shift of the position of the Bragg peak ($\Delta z$) measured in water, when the beam is passing through the material probe of thickness $l$,

$$\text{rWEPL} = \frac{\Delta z}{l}. \qquad (6)$$

$\Delta z$ was determined as the averaged measurements at the 90% distal and 90% proximal dose levels of the Bragg peak. Beam energies equal to 148.80 MeV and 279.97 MeV/u, for pro- tons and carbon ions, respectively, were adopted in WEPL measurements since these energies are approximately at the halfway point of the commissioned energy ranges.

For titanium, a rWEPL of 3.13 was adopted, as reported in Jäkel[31] corresponding to 3080 HU which is the upper limit of the HU scale for this scanner. The CT-HU calibration curves were then implemented into the TPS for both particle types and scan protocols used for treatment planning. In Fig. 3, the calibration curves for carbon ions are shown. The calibration curve should not be used for materials with values higher than 1200 HU (essentially metal implants) unless the WEPL is not actually measured as for the titanium.

For those cases, a specific material override should be used for planning.[3] Calibration curves were then validated by measuring the dose distributions under inhomogeneous conditions, using the CIRS 062 electron density phantom fixed to the beam entrance wall of the PTW MP3-P water phantom (Fig. 4). For each scan protocol used, plans were calculated for a cylindrical target volume contoured in the water phantom. The plans were then delivered and the doses measured with PTW PinPoint ionization chambers placed in water[9] behind each of the tissue inhomogeneities, at the center of the SOBP.

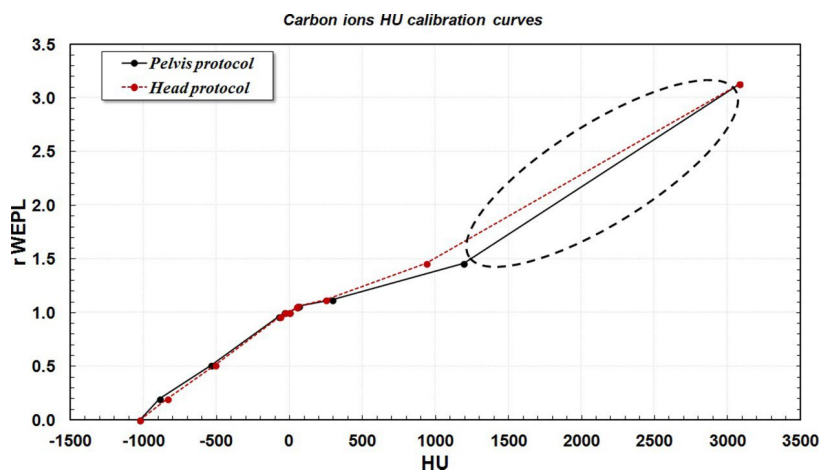

Figure. 3. CT-HU calibration curve for two adopted CT-scan protocols: examples for carbon ions. rWEPL is the relative water-equivalent path length, HU is the Hounsfield unit. The HU values may depend on the CT-protocol adopted such as the case for dense bone tissue, around 1000 HU. The ellipse indicates the region where the curve should not be used (see text for details).

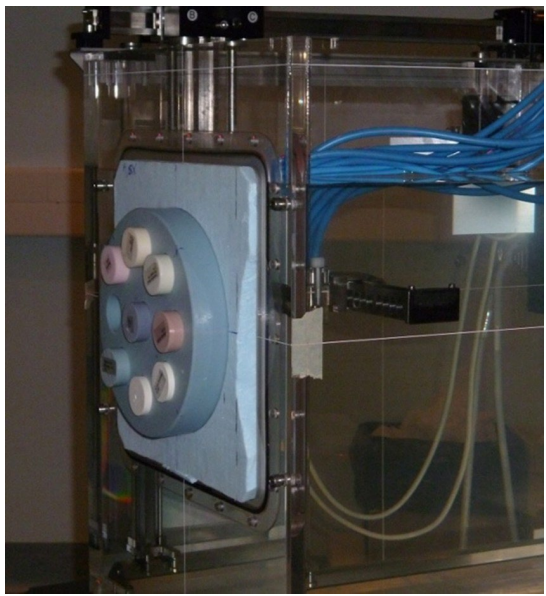

Figure. 4. Inhomogeneity conditions experimental setup: the CIRS 062M phan- tom is positioned in front of the PTW MP3-P water phantom entrance window.

## 2.B  Daily quality assurance

Table II summarizes the main periodic QA procedures adopted at CNAO, together with their frequencies and tolerances, which are in agreement with the relevant literature.[11–15] For daily QA, a subset of ten representative energies, for both particle types, were chosen as a reference for every beam line. In order to verify the whole range approximately every two weeks, one selected energy is analyzed daily during QA. Most of the tests performed during commissioning are being routinely adopted in daily QA for each room and particle type. Tests of dose homogeneity in 2D-fields, spot size, spot position over the scan field, and BM calibration curve are performed as previously described.

In the following paragraphs, the daily performed tests are described, focusing specifically on test procedures differing from those performed during commissioning.

Table II. A list of QA procedures adopted at CNAO with their related frequencies and tolerances.

| Test type | Beam type | Method | Tolerance | Frequency |
|---|---|---|---|---|
| Energy constancy (cf. Sec. 2.B.1) | Monoenergetic pencil beam ($p$)/uniform scanning beam, monoenergetic (C-12) | EBT3 films + All-in-one phantom | 1 mm ($p$)/0.6 mm (C-12) | Daily |
| Spot size (cf. Sec. 2.B.2) | Monoenergetic pencil beam | EBT3 films + film holder | 1 mm or 10% | Daily |
| Spot position (isocenter) (cf. Sec. 2.B.3) | Monoenergetic scanned pencil beam | Strip chambers Monitor1 | 1 mm | Daily |
| Spot position over the scan field (cf. Sec. 2.B.3) | Monoenergetic scanned pencil beam | EBT3 films + All-in-one phantom | 1 mm | Daily |
| Dose homogeneity in 2D-fields (cf. Sec. 2.B.4) | Uniform scanning beam, monoenergetic | EBT3 films + film holder | 5% | Daily |
| BM calibration constancy (cf. Sec. 2.B.5) | Uniform scanning beam, monoenergetic | Farmer IC[a] + water phantom | 2% for $k_{fit}$ 3% for $C(E_i)$ | Daily |
| BM reproducibility ($R$) and proportionality ($P$) (cf. Sec. 2.B.5) | Uniform scanning beam, monoenergetic | Farmer IC + water phantom | 1% ($R$) and 2% ($P$) | Weekly ($R$) and monthly ($P$) |
| Dose in a reference volume (cubic volume) (cf. Sec. 2.B.6) | Modulated scanning beam, SOBP | Farmer IC + water phantom | 5% | Daily |
| Dose distributions in water phantom (cf. Sec. 2.B.6) | Modulated scanning beam, SOBP | PinPoint + 3D holder | 5% | Half-yearly |
| Dose distributions in inhomogeneous phantom (cf. Sec. 2.A.9) | Modulated scanning beam, SOBP | PinPoint + inhomogeneous phantom | 5% | Half-yearly |
| Energy constancy (cf. Sec. 2.A.1) | Monoenergetic pencil beam | Peakfinder | 0.5 mm at 90% of distal dose | Half-yearly |
| Beam contamination (cf. Sec. 2.A.2) | Monoenergetic pencil beam | Peakfinder | No contaminants | Yearly |
| CT-range calibration (cf. Sec. 2.A.9) | Monoenergetic pencil beam | CT scan + Peakfinder | 5 $HU_{ref}$[b] or 1.5% of $HU_{ref}$ values for any material | Yearly |

### 2.B.1. Energy constancy

Since measurements with the Peakfinder would be highly time-consuming in a daily QA session, a simpler and faster procedure to test the energy constancy was established during the beam commissioning phase. In this procedure, EBT3 radiochromic films are inserted into a solid RW3 slab phantom and positioned parallel to the beam direction. One energy per day is verified delivering a single spot on the central axis.[32,33] The measured WEPL for the RW3 is equal to 1.048; therefore, the Bragg peak positions measured with EBT3 films are 4.8% shorter than in water. The energy check is passed if the measured position of the Bragg peak differs less than 1 mm with respect to the reference values measured in the same way.

This procedure was validated for protons. For carbon ions, this method was not applicable as the response of EBT3 is highly dependent on the linear energy transfer (LET), i.e., a strong quenching effect occurs at the end of carbon ions range where LET is very high and the peak cannot be accurately determined.[32,33] Therefore, for carbon ions, the following procedure was established: a monoenergetic uniform rectangular field (3 ×13 cm² size) is delivered to an EBT3 film which is placed behind the dedicated double-wedge of the PTW All-in-one phantom (Fig. 2). For beam energies higher than 297.97 MeV/u, a 10 cm thick RW3 slab phantom is additionally positioned in front of the phantom wedges, acting as a range shifter for practical reasons. The blackening of the EBT3 depends on the wedge thickness traversed by the beam. From film analysis, a relationship between the measured field size in the horizontal plane and the particle energy was found during commissioning of the beam.

A variation of 2 mm of the Bragg peak position in water was seen to correspond to a change in field size by 1.2 mm. To maintain the results within the tolerance level of 1 mm in water for carbon ions, the energy check is passed if the measured field size differs by less than 0.6 mm with respect to the reference values.

### 2.B.2. Spot size in air

The test is performed as described in Sec. 2.A.3. As de- scribed by Chanrion *et al.*,[34] variation in spot dimension within ±25% deteriorates the dose distribution only in a negligible or moderate way. As a consequence, taking into account the typical spot size, the tolerance for this check was established to be ±1 mm or ± 10% (whichever is greater) from the reference values acquired during commissioning.

### 2.B.3. Spot position at the isocenter and across the scan field

To verify the correct position of the nondeflected pencil beam (nominal zero position) with respect to the isocenter, a specific test is performed during daily QA. For each beam line and for both particle species available, approximately 30 monoenergetic pencil beams across the whole energy range are delivered. The spot positions are measured using the strip chambers of Monitor1 (Fig. 5). The test is passed if, for all energies, the deviation between the measured and planned position is less than ± 1 mm for both transversal directions.

Spot position is also tested across the scan field. For this test, same procedure described in Subsection 2.A.4 is adopted. The isocenter position and eight off-axis spot positions are tested. The test is passed if the deviation from the nominal position is within ±1 mm.

### 2.B.4. Dose homogeneity in 2D-fields

The 2D-homogeneity is evaluated for square monoenergetic uniform fields (6× 6 cm$^2$) in the 70% inner portion of the field for transversal profiles along main axes and diagonals. One energy per day is analyzed and the test is passed if $H \leq 5\%$. On a monthly basis, larger monoenergetic squared fields of 15× 15 cm$^2$ are tested and the same tolerance is applied.

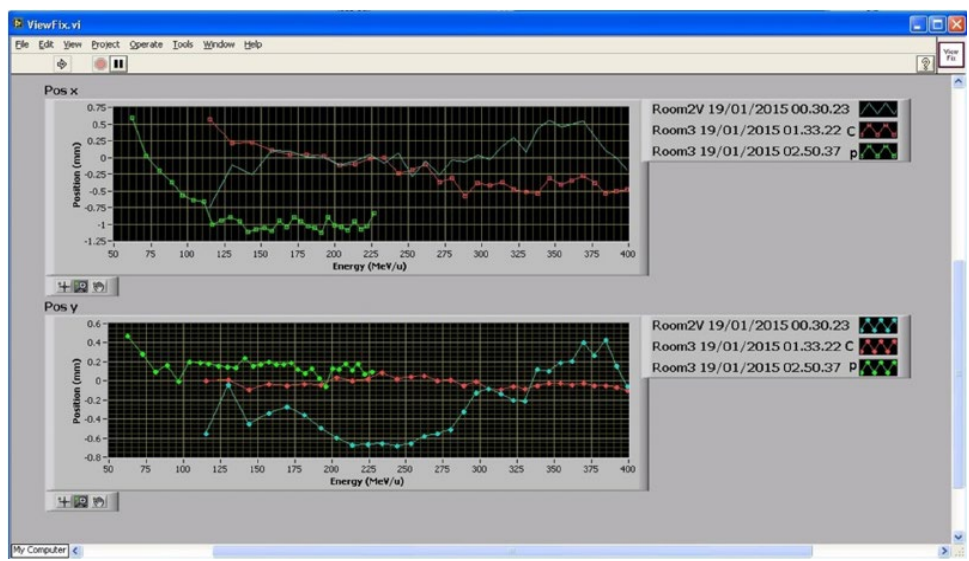

Figure 5. Beam position measured using the strip chambers of Monitor1: example of deviation in the two directions orthogonal to the beam propagation axis for three different beam lines. Beam deviation from the (0;0) position is within ±1

### 2.B.5. Absorbed dose to water under reference conditions and BM calibration check

The BM calibration is checked daily by measuring the calibration factors at a specific energy $E_i$, $C(E_i)$. In the first year of clinical activity, several beam energies had been evaluated daily for every beam line. The calibration curve shape was found to be very stable and therefore, the number of checked energies was initially reduced to two and, then, to one. The tolerance level for each single coefficient $C(E_i)$ was set to 3% with respect to the reference value, while the tolerance on the mean of the calibration factors, $k_{fit}$, was set to 2%. At present, since only one energy is measured, $C(E_i)$ $k_{fit}$, the tolerance level of 2% is applied.
Based on daily measurements, BM calibration curve can be changed, if required, by acting on the scaling factor $k_{fit}$.
A check of the short-term BM reproducibility is performed on a weekly basis. Five consecutive measurements of $C(E_i)$ are collected and the test is passed if the relative standard deviation is within 1%. The BM proportionality is checked monthly by delivering four different dose levels (approximately 0.5, 1, 2, and 5 Gy). For this test, 2% tolerance level is set.

### 2.B.6. Dose in a SOBP

One of the cubes defined in Subsection 2.A.8 is delivered on a daily basis and the dose at the center of the cube is measured with a Farmer chamber. The measured dose is then compared to the reference value calculated by the TPS, using the following formula:

$$d = 100 \times \frac{|D_{meas} - D_{ref}|}{D_{ref}}. \qquad (7)$$

The applied tolerance is 5%.

### 2.B.7. Calibration table for CT scanner

The constancy of HU numbers for water is checked during CT QA protocol on a daily basis, while the existing calibration curves are checked yearly.
Deviations of the HU numbers for each material from the reference value are accepted if the equivalent shift in water of the Bragg peak is within 0.5 mm when crossing 10 cm of any material. Therefore, a tolerance level of 5 HU or 1.5% (whichever is greater) of the reference HU value is defined, for any material.
ions and protons, respectively. A momentum spread of 0.10% and 0.15% for carbon ions and protons, respectively, was used.
The widths of the Bragg peaks in water, in terms of FWHM, are reported in Table III.
As a result of these findings, ripple filters are not used for protons but two of them are used for carbon ions in clinical practice.

### 2.B.8. Documentation

In order to properly record and document all the data regarding machine stability assessed during periodical QA, the irradiated films are date stamped and permanently stored. They are also saved in an electronic format once scanned. The results on dose measurements are also saved in an Excel-spreadsheet and summarized on a paper form signed by a medical physicist, then classified and stored by date, treatment room, and particle species.

http://dx.doi.org/10.1118/1.4928397

## 3 RESULTS

### 3.A Beam commissioning results

#### 3.A.1 Depth-dose distributions

Experimental IDD curves measured using the Peakfinder system were highly reproducible. For both particle species, the deviation between measured and MC simulated Bragg peak positions was within 0.15 mm for protons and within 0.10 mm for carbon ions. In Fig. 6, the comparisons between the experimental data and FLUKA simulated IDDs in water are presented, for protons with a single 2 mm thickness ripple filter and for carbon ions with a pair of ripple filters, each of 2 mm thickness.

In Fig. 7, the percentage differences between FLUKA deposited energy scored, taking into account the Bragg peak chamber area, and a larger area ($25 \times 25$ cm$^2$) for IDD calculations, are shown.
Fine tuning of the main physical parameters (such as beam momentum spread and ionization potential of water) leads to an excellent agreement between experimental and simulated IDDs.[9,21] In particular, ionization values of 77.5 and 77.0 eV were found to closely reproduce the measurements for carbon ions and protons, respectively. A momentum spread of 0.10% and 0.15% for carbon ions and protons, respectively, was used. The widths of the Bragg peaks in water, in terms of FWHM, are reported in Table III. As a result of these findings, ripple filters are not used for protons but two of them are used for carbon ions in clinical practice.

#### 3.A.2 Spot size in air

The FWHM of the pencil beam did not depend on the specific beam line or treatment room. Only slight differences, within 0.5 mm, were shown for the carbon ions vertical line in comparison with the horizontal one A notable widening of the beam spot as a function of air distance from the nozzle (Monitor2) was found for protons [Fig. 8(a)]. As expected, the FWHM was energy-dependent and varied, at the isocenter, from 2.2 to 0.7 cm. The equation reported in Fig. 8(b) closely fits the results.

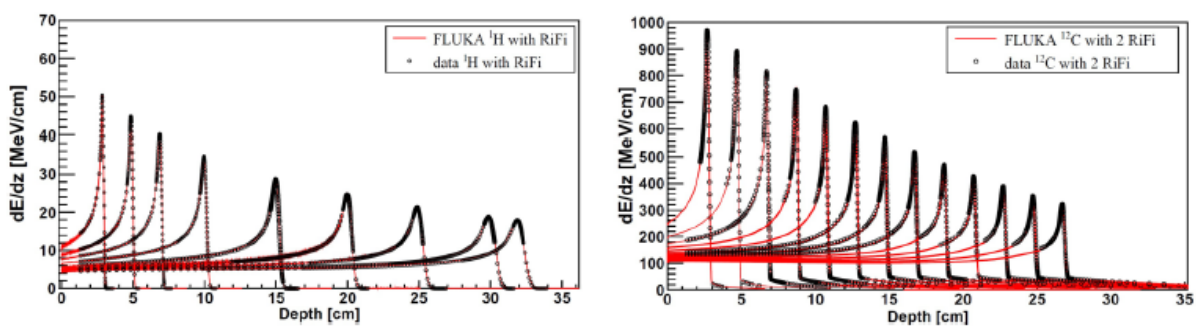

Figura 6. Comparison between FLUKA calculated Bragg peaks in water (lines) for proton with a single 2 mm ripple filter (left panel) and for carbon ions with two 2 mm ripple filters (right panel) and experimental data (points).

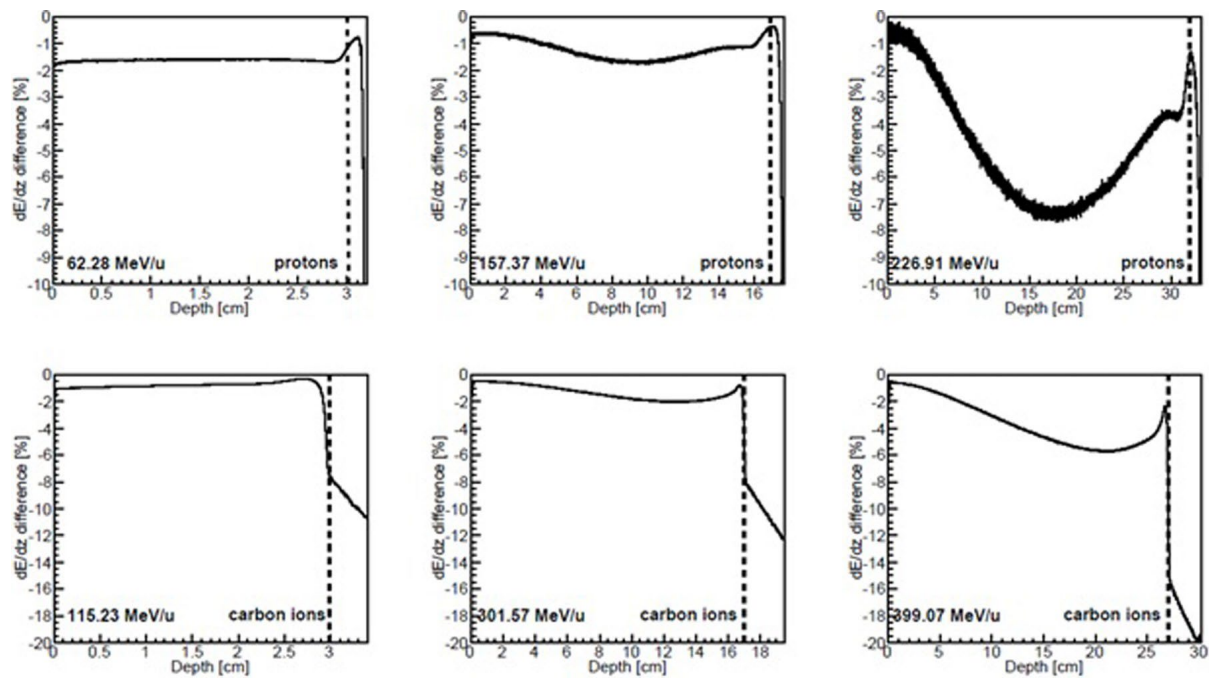

Figure. 7. Percentage differences between FLUKA deposited energy scored taking into account the Bragg peak chamber area and a much larger area (25 25 cm$^2$) for IDD calculations for protons without ripple filter (upper panels) and for carbon ions with two 2 mm ripple filters (lower panels) as a function of depth in water for three different beam energies. A dashed line marks the nominal Bragg peak position for the specified energy.

The in-air scattering is more pronounced for protons (espe- cially at lower energies) than for carbon ions [Fig. 8(c)]. More- over, the spot size at the isocenter in air for carbon ions (smaller focus) varied with energy from 0.8 to 0.4 cm without a ripple filter and from 1.0 to 0.5 cm with two ripple filters [Fig. 8(d)].

The dependence of spot size on spot position within the whole scanning area was negligible for both particle types.

### 3.A.3 Absorbed dose to water under reference conditions

The absorbed dose to water, $D_w$, was measured both with the Farmer and the Advanced Markus chamber. Dose values, measured by the former chamber, were found to be in agreement to within 1% when compared to the Advanced Markus chamber, taken as a reference. The same level of agreement between chambers was found in dose measurements in the center of the SOBPs.

Dose proportionality and reproducibility, as well as beam intensity dependence, were checked for each BM during the commissioning. The proportionality was found to be within 1% in the range of interest (from 0.5 to 5 Gy), short-term reproducibility showed a relative standard deviation of 0.3%–0.5%, and the beam intensity dependence was within 1%.

The agreement between TPS-calculated and measured doses in homogeneous conditions, assessed with Farmer and Advanced Markus chambers, was evaluated during commissioning. The results agreed within 2% for each of the cubic volumes. These findings demonstrated the accuracy of TPS calculation in water. So, in this phase, the reference values for QA (Subsection 3.B.2) were established for the doses in the center of SOBPs.

Table III. The widths of the Bragg peaks in water, in terms of FWHM, are reported for both particle species for the minimum and maximum available energies, with and without ripple filters (RiFi). Their relative distances from the isocenter are also reported in brackets.

| Particle | Energy (range in water) | 0 RiFi (mm) | 1 RiFi (680 mm) (mm) | 2 RiFis (680 and 980 mm) (mm) |
|---|---|---|---|---|
| $p$ | 62.73 MeV (30 mm) | 2.8 | 4.8 | — |
| C-12 | 115.23 MeV/u (30 mm) | 0.9 | — | 5.9 |
| P | 228.57 MeV (320 mm) | 26.1 | 26.2 | — |
| C-12 | 398.84 MeV/u (270 mm) | 9.3 | — | 15.5 |

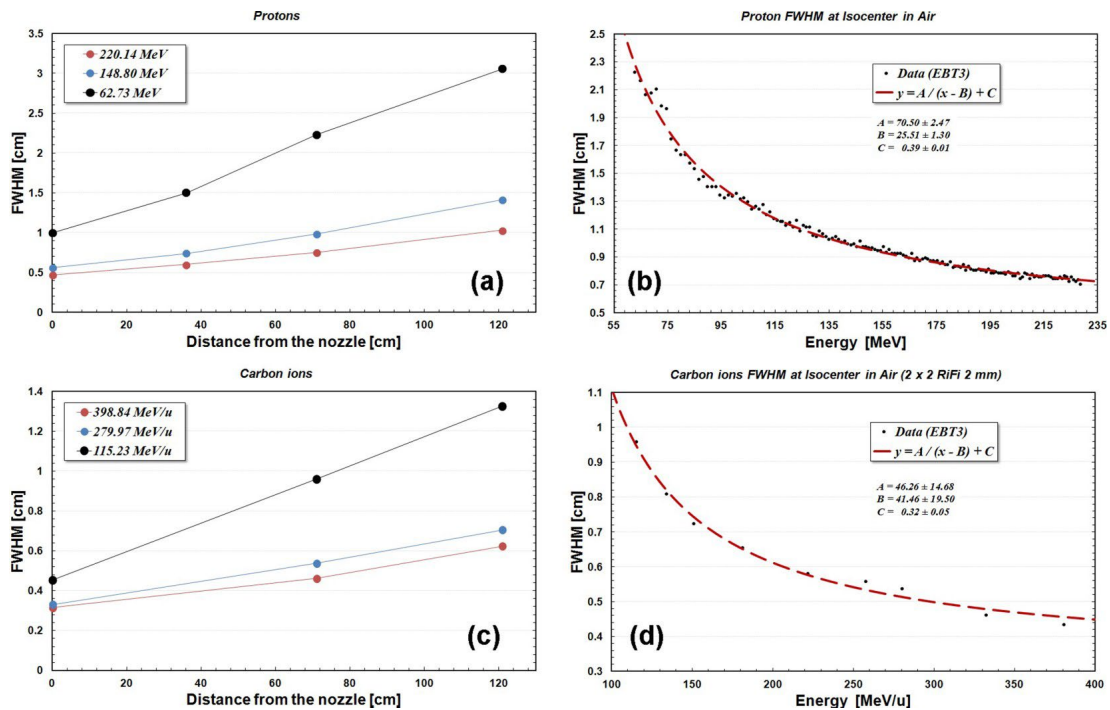

Figure. 8. Beam widening in air for three representing proton energies (a) and proton spot size at the isocenter as a function of energy (b). Beam widening in air for three representing carbon ion energies (c) and carbon ion spot size at the isocenter as a function of beam energy (d).

### 3.A.4 Dosimetric validation of the CT calibration curve

The CT calibration curves were validated measuring dose distributions under inhomogeneous conditions as described in Subsection 2.A.9.

For protons, a close agreement between the measured dose behind each of the tissue inhomogeneities of the CIRS phantom and the TPS-calculated dose was found. Taking into account all tissue inhomogeneities, the mean percent deviation from the reference dose values was 1.35%, ranging from 4.0% to 0.2%.

Similar results were also found for carbon ions, for which the mean percent deviation from the reference dose values was 0.75%, ranging from 4.3% to 0.1%.

### 3.B QA results

In this section, results of 24 months of daily QA checks will be shown. For each of the daily tests, a set of approximately 700 measurements was performed and analyzed.

### *3.B.6 Energy constancy*

For the energy checks, no appreciable deviation of the beam range from reference values was found for both protons and carbon ions. The mean deviation of the absolute value of the difference between measurements and reference values was (0.2 0.1) mm for carbon ions and (0.4 0.4) mm for protons. The corresponding mean percent deviation was 0.07% and 0.14%, respectively. No dependence on beam line or energy was found.

### *3.B.7 BM calibration and dose in SOBP*

Within the whole period under investigation, only minor changes of approximately 1%–2% were globally applied to the BM calibration curves, $F(E)$ s, thus confirming the reliability and stability of the BM. The dose measured in the center of SOBPs (see Subsection 2.A.8) was also found to be very stable for each of the four dose cubes. For the duration of the reviewed period, the measurements performed with the Farmer chamber showed that the mean of the ratios between measured and reference dose $D_{meas}/D_{ref}$ was equal to 0.996 0.013 and 0.987 0.011 for carbon ions and protons, respectively. Additionally, 100% of daily tests were found to be within the defined tolerances.

### *3.B.8 Spot size*

The spot size (in terms of FWHM) was found to be constant for both particle types, all energies, and beam lines. Additionally, the 98.9% of the measurements performed in daily QA sessions were within the tolerance. For carbon ions, the absolute value of the difference between the measured spot sizes and the reference values showed a mean deviation of 0.4 ±0.3 mm while for protons, it was 0.1±0.2 mm. Over the reviewed period, the overall mean spot size deviation from the reference values (averaged for all energies) was approximately 6.7% and 4.1% for carbon ions and protons, respectively.

### *3.B.9 Dose homogeneity in 2D-fields*

The observed dose homogeneity of the scanning fields for all energies, particle types, and beam lines was within the tolerance in 95% of cases, thus demonstrating the excellent behavior and stability of the beam scanning and beam monitor systems. In the analyzed period, mean values of (4.4± 0.9)% for carbon ions and (4.0± 0.9)% for protons were obtained.

The higher energy carbon ion beams generally showed a worse homogeneity level (i.e., higher flatness index) due to the smaller spot size, the scanning step being set at 2 mm for all energies.

### *3.B.10  Spot position accuracy*

Spot position at the isocenter and over the scan field showed a mean deviation within 1 mm. For the 98.6% of the daily sessions analyzed, the spot positions were within tolerance both in the center and at the edges, for all particles, energies, and beam lines.

## 4 DISCUSSION

The commissioning of particle beams has been successfully carried out and daily tests were analyzed over a period of more than 24 months. CNAO proton beam characteristics are also comparable with data previously published for protons;[12] in particular, the strong dependence of spot size at the isocenter on beam energy, as well as the effect of the finite size of Bragg peak chambers on the acquired IDDs, is similar in the two studies.

Due to the results obtained during the geometric and dosimetric characterization of the beams, a transversal scanning step of 2 mm for carbon ions and 3 mm for protons is currently used in clinical practice at CNAO. The spacing of the available energies is generally set to 2 mm for both particle types.

Dose determination in case of proton and carbon ion beams has been previously described in Sec. 2.A.6. Nevertheless, in case of carbon ions, a deeper evaluation of some of the uncertainties involved in dose calculation needs to be discussed.

First of all, as reported in Karger et al.,[14] many different $W$-values have been reported in the literature. Two independent studies published by Brede et al.[35] and Sakama et al.[36] showed an underestimation of approximately 3%–3.5% in the determination of $D_w$ for carbon ion beams when ion chambers are used. In the first case, a significant deviation between ionization chamber dosimetry and calorimetry was only found for carbon ions (not in case of protons, neutrons, or photons); in the second case, a $W$-value in air for carbon ion beams equal to 35.72% 1.5% J C$^{-1}$ was found, which is 3.5% higher than the value suggested by IAEA (34.50 J C$^{-1}$) and used for $k_Q$ determination. Further studies are needed to confirm or deny these findings.

In addition, a number of studies[37–39] have showed that the value of the stopping-power ratios depends on the beam energy and depth in water, thus slightly deviating from the constant 1.130 value adopted by IAEA.

In conclusion, taking into account the aforementioned uncertainties, and the relatively small correction due to the ion recombination ($k_s$ ranges from 1.005 to 1.008), we decided to prudentially apply a minor deviation from IAEA dosimetry protocol for carbon ions. A comprehensive multiplying term $k_{C-12}$ equal to 1.03 (that is a correction similar to the relative standard uncertainty for cylindric chambers and carbon ions estimated by IAEA) was applied for carbon ions to the calculation of $D_w$.

Our QA protocol appears consistent with the published literature in terms of check procedures, frequencies, and applied tolerances.[11,42] Moreover, the main treatment parameters, such as delivered number of particles per spot, spot position accuracy, presence/absence, and the position of passive elements (range shifters and ripple filters), as well as other fundamental conditions for successful dose delivering, like the correct functioning of the field programmable gate arrays (FPGAs) managing scanning magnets, strip, pixel, and transmission chambers, are continuously monitored by the BM connected to the so-called patient interlocks system (PIS). If any tolerance level is exceeded, the BM generates an interlock and the irradiation is immediately stopped.[5] The PIS is also connected to PPS to prevent irradiation unless the couch is in a locked state. Full functionality tests of the PIS system are routinely performed during the half-yearly scheduled preventive maintenance.

So far, all the three treatment rooms are in clinical operation: three horizontal carbon ion beam lines, one vertical carbon ion beam line, and two horizontal proton beam lines have been successfully commissioned. The commissioning of the remaining two beam lines for protons is scheduled for 2015.

Dosimetric tests aiming at evaluating the suitability of an array of microdiamond detectors for patient-specific plan verification are also ongoing.[40]

The results of beam commissioning and daily QA procedures reported here were included in the technical documentation submitted in 2011 by CNAO to the Italian Ministry of Health[41] for the authorization request to start clinical activity under experimental phase regimen. The CNAO treatment de- vice, including the synchrotron and all the related subsystems, achieved the CE mark as a medical device in December 2012. The first patient was treated with proton beams in September 2011, while the first carbon ion treatments started in November 2012. As of October 2014, approximately 350 patients have been treated.

Based on the experience gained during the first year of activity, the following action level has been set for daily checks involving the use of films: if the result of a test exceeds twice the tolerance level or it remains out of tolerance for two consecutive days, patient treatments are temporarily stopped until the problem is solved. The first scenario has not happened so far, while the second one was extremely rare (ten cases out of approximately 800 days of treatment). In those cases, the cause of the deviation was always related to the beam optics and required beam steering adjustment. Any sporadic results slightly out of tolerance for 1 day were never regarded as critical and could mostly be due to undetected experimental errors, such as film misalignment.

Concerning the feasibility of our daily beam QA protocol, measurements are performed during every night by a medical physicist or a well-trained radiation technologist and approximately 60 min per beam line and particle type are needed. Any remaining time in the night is allocated for weekly, monthly, half-yearly, and yearly QA or patient-specific dosimetric QA tests. The night shift lasts 8 h from 10:00 p.m. to 6:00 a.m., from Sunday night to Friday morning. The EBT3 films irradiated during QA are then scanned and analyzed early in the morning (around 7:30 a.m.) by the medical physicist on duty.

The standard beam time schedule at CNAO foresees, from Monday to Friday, around 14 h of patient treatments (from 8:00 a.m. to 10:00 p.m.). Moreover, two additional QA shifts (8 h each) are allocated to the medical physics staff dur- ing weekends for patient-specific QA checks and research activities.

From November 2011 to August 2014, about 800 days of treatments were provided at CNAO. During this period, there were 42 days of scheduled preventive maintenance. Usually a maintenance period, consisting of five consecutive days including the weekend, is foreseen every 3 months. Short maintenance activities are also scheduled during the week- ends, together with machine development and research. Therefore, approximately 4.5% machine downtime is due to planned maintenance activity. Thirteen treatment days were lost due to machine breakdowns, representing about 1.5% of the down- time.

In these instances, patient treatments were never stopped for more than 1 day at a time. The main causes of the failures were related to faults of power supply components, linear accelerator, ion sources, infrastructure network, water cooling, or vacuum systems. Excluding the mandatory scheduled preventive maintenance of the synchrotron, this relatively low rate of downtime due to machine failures confirms the reliability of the machine itself. This can be considered to be a remarkable result, since CNAO is a fully in-house-designed and developed high-technology project.

## 5 CONCLUSIONS

The physical commissioning of proton and carbon ion beams delivered with active pencil beam scanning technique has been reported in this paper. The large number of measure- ments performed during the first 24 months of tests has shown the stability of the main beam characteristics and the overall reliability of the CNAO system for clinical use.

The results of this study may be especially useful for med- ical physicists working in new light ion therapy facilities in setting up a program for beam commissioning and periodic QA checks.

## ACKNOWLEDGMENTS

The authors would like to thank Dr. Eleanor Dyke and Darren Thompson (Medical Physics, Oxford University Hospital—Newcastle upon Tyne Hospitals NHS Foundation Trust, UK) for helpful discussion and his valuable contribution in editing the paper.

[a)]Author to whom correspondence should be addressed. Electronic mail: mirandola@cnao.it; Telephone: (+39) 0382 078511.

## REFERENCES

[1]http://www.ptcog.ch/index.php/facilities-under-construction last visited: January 2015.
[2]M. Engelsman, M. Schwarz, and L. Dong, "Physics controversies in proton therapy," Semin. Radiat. Oncol. **23**(2), 88–96 (2013).
[3]M. C. Cantone, M. Ciocca, F. Dionisi, P. Fossati, S. Lorentini, M. Krengli,S. Molinelli, R. Orecchia, M. Schwarz, and I. Veronese, "Application of failure mode and effects analysis to treatment planning in scanned proton beam radiotherapy," J. Radiat. Oncol. **8**(1), 127 (2013).
[4]S. Rossi, "The National Center for Oncological Hadrontherapy (CNAO): Status and perspectives," Phys. Med. **31**(4), 333–351 (2015).
[5]S. Giordanengo, M. A. Garella, F. Marchetto, F. Bourhaleb, M. Ciocca, A. Mirandola, V. Monaco, M. A. Hosseini, C. Peroni, R. Sacchi, R. Cirio, and M. Donetti, "The CNAO dose delivery system for modulated scanning ion beam radiotherapy," Med. Phys. **42**, 263–275 (2015).
[6]A. Pella, M. Riboldi, B. Tagaste, D. Bianculli, M. Desplanques, G. Fontana, P. Cerveri, M. Seregni, G. Fattori, and R. Orecchia, "Commissioning and quality assurance of an integrated system for patient positioning and setup verification in particle therapy," Technol. Cancer Res. Treat. **13**, 303–314 (2014).
[7]M. Desplanques, B. Tagaste, G. Fontana, A. Pella, M. Riboldi, G. Fattori, A. Donno, G. Baroni, and R. Orecchia, "A comparative study between the imaging system and the optical tracking system in proton therapy at CNAO," J. Radiat. Res. **54**(1), i129–i135 (2013).
[8]A. Facoetti, B. Vischioni, M. Ciocca, M. Ferrarini, Y. Furusawa, A. Mairani, Y. Matsumoto, A. Mirandola, S. Molinelli, A. Uzawa, G. Vilches Freixas, and R. Orecchia, "*In vivo* radiobiological assessment of the new clinical carbon ion beams at CNAO," Radiat. Prot. Dosim. (2015) [E-pub ahead of print].
[9]S. Molinelli, A. Mairani, A. Mirandola, G. Vilches Freixas, T. Tessonnier, S. Giordanengo, K. Parodi, M. Ciocca, and R. Orecchia, "Dosimetric accuracy assessment of a treatment plan verification system for scanned proton beam radiotherapy: One-year experimental results and Monte Carlo analysis of the involved uncertainties," Phys. Med. Biol. **58**(11), 3837–3847 (2013).
[10]International Commission on Radiation Units and Measurements, "ICRU report 78," J. ICRU **7**(2), 135–139 (2007).
[11]C. P. Karger, G. H. Hartmann, O. Jäkel, and P. Heeg, "Quality management of medical physics issues at the German heavy ion therapy project," Med. Phys. **27**(4), 725–736 (2000).
[12]M. T. Gillin, N. Sahoo, M. Bues, G. Ciangaru, G. Sawakuchi, F. Poenisch, B. Arjomandy, C. Martin, U. Titt, and K. Suzuki, "Commissioning of the discrete spot scanning proton beam delivery system at the University of Texas MD Anderson Cancer Center, Proton Therapy Center, Houston," Med. Phys. **37**(1), 154–163 (2010).
[13]G. H. Hartmann, O. Jäkel, P. Heeg, C. P. Karger, and A. Kriessbach, "Determination of water absorbed dose in a carbon ion beam using thimble ionization chambers," Phys. Med. Biol. **44**(5), 1193–1206 (1999).
[14]C. P. Karger, O. Jäkel, H. Palmans, and T. Kanai, "Dosimetry for ion beam radiotherapy," Phys. Med. Biol. **55**(21), R193–R234 (2010).
[15]O. Jäkel, G. H. Hartmann, C. P. Karger, P. Heeg, and S. Vatnitsky, "A calibration procedure for beam monitors in a scanned beam of heavy charged particles," Med. Phys. **31**(5), 1009–1013 (2004).
[16]*syngo* RT Planning, Physics Reference for Particle Treatment Planning, T11–050.640.05.02.02, Siemens AG, Germany, 2010.
[17]T. T. Böhlen, F. Cerutti, M. P. W. Chin, A. Fassò, A. Ferrari, P. G. Ortega, A. Mairani, P. R. Sala, G. Smirnov, and V. Vlachoudis, "The FLUKA Code: Developments and challenges for high energy and medical applications," Nucl. Data Sheets **120**, 211–214 (2014).
[18]A. Ferrari, P. R. Sala, A. Fassò, and J. Ranft, "FLUKA: A multi-particletransport code," INFN-TC-05–11, SLAC-R-773, CERN-2005-10, 2005.
[19]A. Mairani, T. T. Böhlen, A. Schiavi, T. Tessonnier, S. Molinelli, S. Brons, G. Battistoni, K. Parodi, and V. Patera, "A Monte Carlo-based treatment planning tool for proton therapy," Phys. Med. Biol. **58**(8), 2471–2490 (2013).
[20]T. Tessonnier, A. Mairani, F. Cappucci, A. Mirandola, G. Vilches freixas, S. Molinelli, M. Donetti, and M. Ciocca, "Development and application of tools for Monte Carlo based simulations in a particle beam radiotherapy facility," Appl. Radiat. Isot. **83**, 155–158 (2014).
[21]K. Parodi, A. Mairani, S. Brons, B. G. Hasch, F. Sommerer, J. Naumann, O. Jäkel, T. Haberer, and J. Debus, "Monte Carlo simulations to support start-up and treatment planning of scanned proton and carbon ion therapy at a synchrotron-based facility," Phys. Med. Biol. **57**(12), 3759–3784 (2012).

[22] F. Bourhaleb, A. Attili, R. Cirio, P. Cirrone, F. Marchetto, M. Donetti, M. A. Garella, S. Giordanengo, N. Givehchi, and S. Iliescu, "Monte Carlo simulations of ripple filters designed for proton and carbon ion beams in hadrontherapy with active scanning technique," J. Phys.: Conf. Ser. **102**(1), 012002 (2008).

[23] L. Lin, C. G. Ainsley, T. Mertens, O. De Wilde, P. T. Talla, and J. E. McDonough, "A novel technique for measuring the low-dose envelope of pencil-beam scanning spot profiles," Phys. Med. Biol. **58**(12), N171–N180 (2013).

[24] L. Lin, C. G. Ainsley, and J. E. McDonough, "Experimental characterization of two-dimensional pencil beam scanning proton spot profiles," Phys. Med. Biol. **58**(17), 6193–6204 (2013).

[25] IAEA, "Absorbed dose determination in external beam radiotherapy: An international code of practice for dosimetry based on standards of absorbed dose to water," FLUKA: A Multi-Particle Transport Code (International Atomic Energy Agency, Vienna, 2000), Technical Report Series 398.

[26] O. Jäkel, G. H. Hartmann, P. Heeg, and D. Schardt, "Effective point of measurement of cylindrical ionization chambers for heavy charged particles," Phys. Med. Biol. **45**(3), 599–607 (2000).

[27] H. Palmans and F. Verhaegen, "On the effective point of measurement of cylindrical ionization chambers for proton beams and other heavy charged particle beams," Phys. Med. Biol. **45**(8), L20–L22 (2000).

[28] M. S. Weinhous and J. A. Meli, "Determining P-ion, the correction factor for recombination losses in an ionization chamber," Med. Phys. **11**(6), 846–849 (1984).

[29] S. Lorin, E. Grusell, N. Tilly, J. Medin, P. Kimstrand, and B. Glimelius, "Reference dosimetry in a scanned pulsed proton beam using ionisation chambers and a Faraday cup," Phys. Med. Biol. **53**(13), 3519–3529 (2008).

[30] O. Jäkel, C. Jacob, D. Schardt, C. P. Karger, and G. H. Hartmann, "Relation between carbon ion ranges and x-ray CT numbers," Med. Phys. **28**(4), 701–703 (2001).

[31] O. Jäkel, "Ranges of ions in metals for use in particle treatment planning," Phys. Med. Biol. **51**(9), N173–N177 (2006).

[32] L. Zhao and I. J. Das, "Gafchromic EBT film dosimetry in proton beams," Phys. Med. Biol. **55**(10), N291–N301 (2010).

[33] M. Martisikova and O. Jäkel, "Dosimetric properties of Gafchromic® EBT films in monoenergetic medical ion beams," Phys. Med. Biol. **55**(13), 3741–3751 (2010).

[34] M. A. Chanrion, F. Ammazzalorso, A. Wittig, R. Engenhart-Cabillic, and U. Jelen, "Dosimetric consequences of pencil beam width variations in scanned beam particle therapy," Phys. Med. Biol. **58**(12), 3979–3993 (2013).

[35] H. J. Brede, K. D. Greif, O. Hecker, P. Heeg, J. Heese, D. T. L. Jones, H. Kluge, and D. Schardt, "Absorbed dose to water determination with ionization chamber dosimetry and calorimetry in restricted neutron, photon, proton and heavy-ion radiation fields," Phys. Med. Biol. **51**(15), 3667–3682 (2006).

[36] M. Sakama, T. Kanai, A. Fukumura, and K. Abe, "Evaluation of w values for carbon beams in air, using a graphite calorimeter," Phys. Med. Biol. **54**(5), 1111–1130 (2009).

[37] O. Geithner, P. Andreo, N. Sobolevsky, G. Hartmann, and O. Jäkel, "Calculation of stopping power ratios for carbon ion dosimetry," Phys. Med. Biol. **51**(9), 2279–2292 (2006).

[38] K. Henkner, N. Bassler, N. Sobolevski, and O. Jäkel, "Monte Carlo simulations on the water-to-air stopping power ratio for carbon ion dosimetry," Med. Phys. **36**(4), 1230–1235 (2009).

[39] D. Sanchez-Parcerisa, A. Gemmel, O. Jäkel, E. Rietzel, and K. Parodi, "Influence of the delta ray production threshold on water-to-air stopping power ratio calculations for carbon ion beam radiotherapy," Phys. Med. Biol. **58**(1), 145–158 (2013).

[40] M. Marinelli, G. Prestopino, C. Verona, G. Verona-Rinati, M. Ciocca, A. Mirandola, A. Mairani, L. Raffaele, and G. Magro, "Dosimetric characterization of a microDiamond detector in clinical scanned carbon ion beams," Med. Phys. **42**, 2085–2093 (2015).

[41] R. Orecchia, V. Vitolo, M. R. Fiore, P. Fossati, A. Iannalfi, B. Vischioni, A. Srivastava, J. Tuan, M. Ciocca, and S. Molinelli, "Proton beam radiotherapy: Report of the first ten patients treated at the Centro Nazionale di Adroterapia Oncologica (CNAO) for skull base and spine tumours," Radiol. Med. **119**(4), 277–282 (2014).

[42] International Commission on Radiation Units and Measurements, "ICRU report 78," J. ICRU **7**(2), 49–81 (2007).